\documentclass[10pt, a4paper]{article}
\usepackage{myStyle1}

\title{EFFECT OF ELECTRON-ELECTRON INTERACTION
NEAR THE METAL-INSULATOR TRANSITION IN DOPED SEMICONDUCTORS
STUDIED WITHIN THE LOCAL DENSITY APPROXIMATION}

\author{YOSUKE HARASHIMA, KEITH SLEVIN}
\address{%
  Department of Physics, Graduate School of Science, Osaka University, \\
  1-1 Machikaneyama, Toyonaka, Osaka 560-0043, Japan \\
  slevin@phys.sci.osaka-u.ac.jp}

\begin{document}

\maketitle

\begin{abstract}

We report a numerical analysis of Anderson localization in a model of a doped semiconductor.
The model incorporates the disorder arising from the random spatial distribution of the
donor impurities
and takes account of the electron-electron interactions between the carriers using
density functional theory in the local density approximation.
Preliminary results suggest that the model exhibits a metal-insulator transition.

\end{abstract}

\section{Introduction}

In semiconductors a zero temperature metal-insulator transition is observed as a function of doping
concentration.
For samples with concentrations below a critical concentration, the conductivity extrapolated to zero temperature is
found to be zero.
For samples with concentrations exceeding this critical concentration, the zero temperature limit of the conductivity
is finite.\cite{Mott1990}
One well studied example is phosphor doped silicon (Si:P) (see Ref. \refcite{Lohneysen2011} for a recent review).
The relative importance of the roles that electron-electron interactions and disorder play in this transition
is still not clear.
The Coulomb interaction between the electrons leads us to expect that the impurity band is split into upper and
lower Hubbard bands and that the transition is associated with a closing of the Hubbard gap.
However, this ignores the effect of the disorder that arises from the random spatial distribution of the donor impurities
and the possibility of Anderson localization.\cite{Anderson1958}
This paper is a preliminary report of numerical simulations designed to address this issue.

\section{Model}
As a simple model of a doped semiconductor we consider an effective medium with electron effective mass $m_{e}^{*}$ and dielectric
constant $\varepsilon_{r}$ equal to those of the host semiconductor crystal.
In this effective medium $N$ donor impurities are randomly distributed in space.
Since we have mind phosphor in silicon, we assume that each donor supplies one electron
and as a result has a net charge of $+e$.
This is the only property of the donor which enters our model.
There are an equal number of electrons so that the total charge is zero.
The electrons interact with the donors through the Coulomb interaction.
The random spatial distribution of the donors thus produces a random potential in which the electrons move.
At the same time the electrons interact with each other via the Coulomb interaction.
The Hamiltonian of this system is
\begin{equation} \label{eq:hamiltonian}
  \mathcal{H}
  = - \frac{1}{2 m_{e}^{*}} \sum_{i=1}^{N} \nabla_{i}^{2}
  - \frac{1}{\varepsilon_{r}} \sum_{i, I=1} ^{N}\;
  \frac{1}{\left| \Vec{r}_{i} - \Vec{R}_{I} \right|}
  + \frac{1}{2 \varepsilon_{r}}
  \sum_{i \neq j} \frac{1}{\left| \Vec{r}_{i} - \Vec{r}_{j} \right|} \; .
\end{equation}
Here,  Hartree atomic units are used.
The positions of the donor impurities are denoted by $\{\Vec{R}_{I}\}$.
The first term is the kinetic energy of the electrons, the second term describes the interaction of the electrons
with the donor impurities, and the third term describes the interaction between the electrons.
A fourth term describing the mutual Coulomb interaction between the donor impurities should also be included
if the correct total energy of the system is required.
However, since the positions of the donor impurities, while random, are fixed,
this contribution to the energy does not play any role in the following discussion and is, therefore, omitted.

To deal with the electron-electron interaction we use density functional theory\cite{Hohenberg1964} and
solve the Kohn-Sham equations\cite{Kohn1965} that describe an auxiliary one-electron problem that has
the same ground state density as the interacting problem of Eq. (\ref{eq:hamiltonian}).
The Kohn-Sham equations are
\begin{equation}
\left(
  - \dfrac{1}{2 m_{e}^{*}} \nabla^{2}
  + V_{\text{eff}} \left[ n \left( \Vec{r} \right) \right]
  \right)
  \phi_{i} \left( \Vec{r} \right)
  = \epsilon_{i} \phi_{i} \left( \Vec{r} \right) \; \; \; \left( i =1,\ldots,N\right) \;,
\end{equation}
where
\begin{equation}\label{eq:veff}
V_{\text{eff}} \left[
   n \left( \Vec{r} \right) \right]
  = - \dfrac{1}{\varepsilon_{r}}
  {\displaystyle \sum_{I=1}^{N}}
  \dfrac{1}{\left| \Vec{r} - \Vec{R}_{I} \right|}
  + \dfrac{1}{\varepsilon_{r}}
  {\displaystyle \int} d^3 r^{\prime} \;
  \dfrac{n \left( \Vec{r}^{\prime} \right)}{\left| \Vec{r} - \Vec{r}^{\prime} \right|}
  + V_{\text{XC}} \left[ n \left( \Vec{r} \right) \right] \;.
\end{equation}
The number density of the electrons is
\begin{equation} \label{ChargeDensity}
n \left( \Vec{r} \right)
= \sum_{i=1}^{N}
\left| \phi_{i} \left( \Vec{r} \right) \right|^{2} \; .
\end{equation}
Periodic boundary conditions are imposed.
In this model the dependence on the medium enters  only through the effective mass and dielectric constant.
Having in mind silicon as the host semiconductor we set
\begin{equation}
    m_{e}^{*} = 0.32 \; ,\; \varepsilon_{r} = 12.0\; .
\end{equation}

The exchange-correlation potential appearing in the Kohn-Sham equations is given by the functional derivative of the exchange-correlation energy
with respect to the number density of electrons
\begin{equation}
    V_{\text{XC}} \left( \Vec{r} \right) = \frac{ \delta E_{\text{XC}}}{\delta n \left( \Vec{r} \right) } \; .
\end{equation}
While in principle the Kohn-Sham equations are exact, in practice the exact form of the
exchange-correlation potential is not known and an approximation is required.
In this work, we use the local density approximation (LDA)
in which the functional is approximated as
\begin{equation}
    E_{\text{XC}} \approx E_{\text{XC}}^{\text{LDA}} \equiv
    \int d^3 r  \epsilon_{\text{XC}} \left( n \left( \Vec{r} \right) \right) n \left( \Vec{r} \right) \;.
\end{equation}
In this preliminary work, we assume complete spin polarization.\footnote{
A calculation that assumes either zero or complete spin polarization is numerically less demanding than a
calculation using the local spin density approximation but excludes the possibility of a Hubbard
gap from consideration.
We do not assume zero polarization because double occupation of the impurities is unlikely in the localized phase.}
We use the form of $\epsilon_{\text{XC}}$ given in Eq. (2) of Ref.~\refcite{Gunnarsson1974} (with spin polarization $\zeta = 1$) though with
the parameter values given in Ref.~\refcite{Janak1975} rather than Ref.~\refcite{Gunnarsson1974}.

In the literature expressions for the
exchange-correlation potential are  given for electrons in free space whereas
we are considering an effective medium.
To map the expressions in the literature to the formulae we require here, we re-scale lengths and energies according to the formulae
\begin{equation}
    \tilde{\Vec{r}} = \left( m_{e}^{*} / \varepsilon_{r} \right) \Vec{r} \; , \; \; \; \;
    \tilde{E} = \left( \varepsilon_{r}^{2} / m_{e}^{*} \right) E \; .
\end{equation}
After this re-scaling we have
\begin{equation} \label{XCPotential}
  V_{\text{XC}} \left[ n \left( \Vec{r} \right) \right]
  = \frac{m_{e}^{*}}{\varepsilon_{r}^{2}} \cdot \tilde{V}_{\text{XC}}
  \left[ \tilde{n} \left( \tilde{\Vec{r}} \right) \right] \; ,
\end{equation}
where
\begin{equation}
    \tilde{n} \left( \tilde{\Vec{r}} \right)
= \left( \varepsilon_{r} / m_{e}^{*} \right)^{3}
n \left( \Vec{r} \right) \; ,
\end{equation}
and $\tilde{V}_{\text{XC}}$ is the exchange-correlation potential found in the literature.

In Hartree atomic units the unit of length is the Bohr
$
    a_0 = \varepsilon_0 h^2 / \left( \pi m_e e^2 \right) \approx 5.292 \times 10^{-11}\text{m}
$.
We use this as the unit of length throughout.
The simulations are performed by generating an ensemble of cubic samples of linear dimension $L$.
In each sample, donor impurities are randomly distributed on a cubic lattice with lattice constant $36$ Bohr.
This prevents impurities being positioned un-physically close together by chance. (We do not allow two
impurities to be at the same site on this lattice.)
The volume of the system is $V=L^3$ and the donor concentration
$
   n_D = N / V
$.

For numerical purposes the continuous description above is replaced by a discrete description
on a real-space grid with spacing $a$.
Derivatives are replaced by next nearest neighbor finite difference approximations.
The resulting matrices and vectors have dimension equal to the number of grid points $\left( L/a \right)^3$.
When calculating the results shown below we used a grid spacing of
\begin{equation}
    a = 18 \; \text{Bohr} \;.
\end{equation}
For comparison the effective Bohr radius for an electron in the conduction band of silicon is
$a_0^{*} = \varepsilon_r/m_e$ Bohr $ \approx 37.5$ Bohr.
The potential due to the positive donor impurities is calculated by expressing the charge density
of the impurities as a Fourier series.
A cut-off is imposed on the wavenumbers so that the number of terms in this series is equal to the number of
points of the real-space grid.
In effect, this replaces the delta-functions of the charge density of the impurities with
an approximate smooth charge density.
Poisson's equation is solved exactly for this approximate density and the corresponding potential obtained using an inverse Fourier transform.
This calculation need only be performed once for a given impurity configuration.
The Hartree like term in Eq. (\ref{eq:veff}) is evaluated in a similar way.
This latter calculation needs to be repeated for each iteration.
Within the LDA the real space finite difference approximation of the Kohn-Sham equations
yields a Hamiltonian matrix that is sparse.
The $N$ Kohn-Sham orbitals of lowest energy are found using the
JADAMILU sparse matrix library.\cite{JADAMILU}
The self consistent solution of the Kohn-Sham equations is found by iteration.

\section{Result}

To determine the nature, localized or extended, of the Kohn-Sham wavefunctions we use multi-fractal finite size scaling
\cite{Rodriguez2010,Rodriguez2011} (MFSS) of
the wavefunction intensity distribution.
The system is divided into boxes of linear dimension $l$ and the Kohn-Sham wavefunction intensity is coarse grained
\begin{equation}
\mu_{k} \equiv \int_{k\text{th box}} d^{3} r \;
\left| \phi_i \left( \Vec{r} \right) \right|^{2} \; .
\end{equation}
From these coarse grained intensities a random variable $\alpha$ is defined
\begin{equation}
    \alpha \equiv \frac{\ln \mu}{\ln \lambda} \; ,
\end{equation}
where $\lambda$ is the ratio of box and system sizes
\begin{equation}
\lambda \equiv \frac{l}{L} \; .
\end{equation}
As  described in Refs.~\refcite{Rodriguez2010} and \refcite{Rodriguez2011}, for Anderson's model of
localization\cite{Anderson1958} the distribution of $\alpha$ is scale invariant at the Anderson transition provided $\lambda$ is held fixed.
The distribution shifts to smaller (larger) values in the metallic (localized) phases as the systems size increases.
We expect similar considerations to apply to the intensities of the Kohn-Sham wavefunctions.
In what follows we focus on the generalized multi-fractal exponent $\tilde{\alpha}_0$ for the highest occupied
Kohn-Sham orbital.
We refer the reader to Eqs. (6), (7) and (19) of Ref.~\refcite{Rodriguez2011} for the definition of this quantity.
The ensemble average in the definition given there is estimated using an average over samples in the usual way.
The precision of this estimate is determined using the formulae given in Table II of Ref. \refcite{Rodriguez2011}.

\begin{figure}[pt]
  \begin{center}
    \includegraphics[width = 10.0cm]{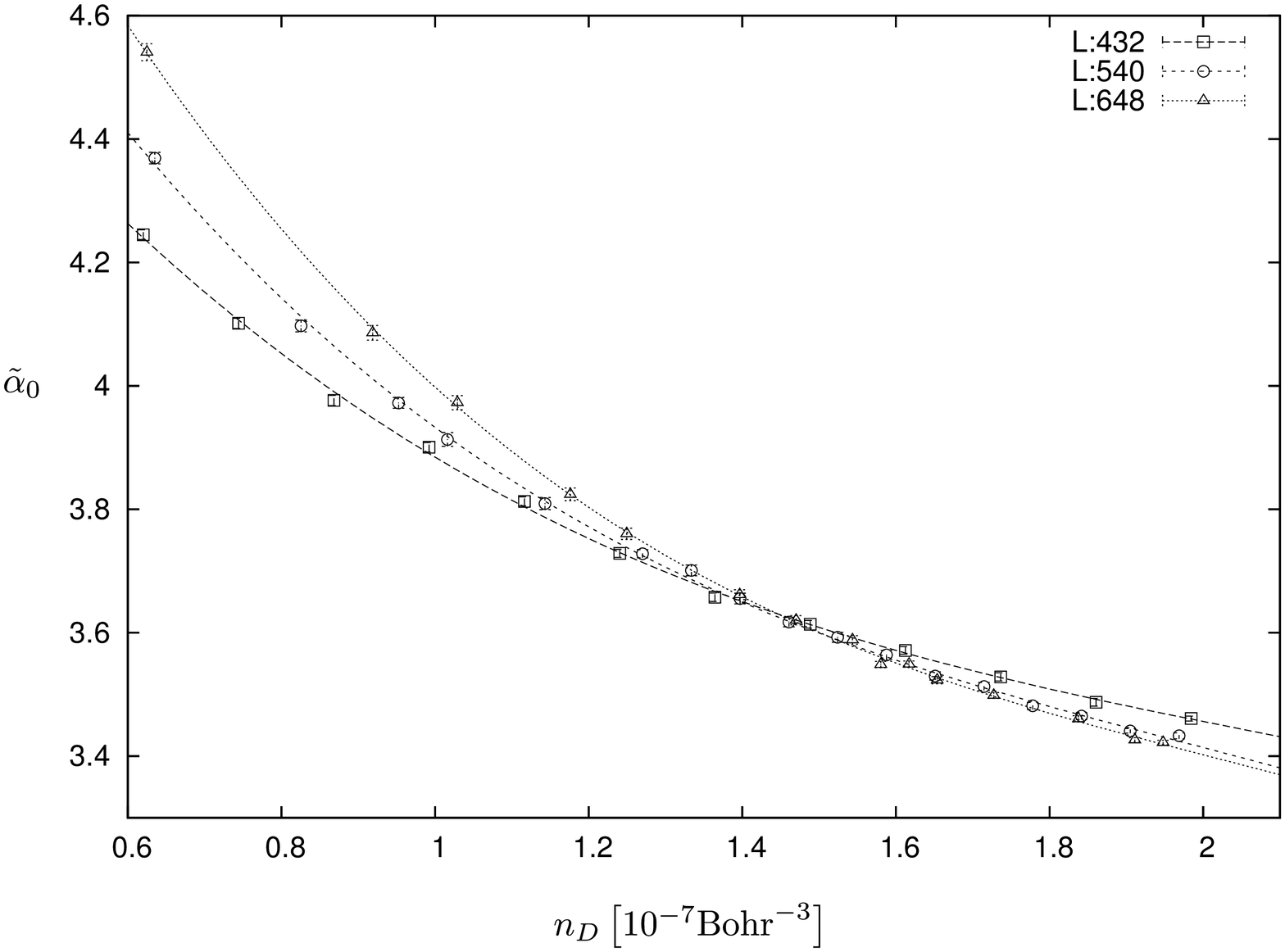}
  \end{center}
  \vspace*{8pt}
  \caption{The  value of generalized multi-fractal exponent $\tilde{\alpha}_0$ as a function of donor impurity concentration
    for $L=432$, $540$, and $648$ Bohr.
    The number of samples simulated varies between 494 and 1004.
    \label{fig1}}
\end{figure}

In Fig.\ref{fig1} we plot the donor concentration dependence of
$\tilde{\alpha}_0$ for three different system sizes.
Least squares fits of third order polynomials have been made to the data for each system size.
For low concentration behavior typical of Anderson localized wavefunctions is seen, i.e. we see
shifts to larger values as the system size increases.
For high concentrations the system size dependence is less pronounced but there seems to be a shift to smaller values with
increasing system size.
(Note there is a lower bound of $\tilde{\alpha}_0\ge 3$ set by wavefunction normalization.)
The polynomial fits show a common crossing point around $1.4 \times 10^{-7}  \text{Bohr}^{-3} $.
The data, though certainly very preliminary, suggest a metal-insulator transition near this value.
For comparison the experimental critical concentration for the metal-insulator transition
in Si:P is $\simeq 5.2 \times 10^{-7}  \text{Bohr}^{-3} $.\cite{Lohneysen2011}
It is possible that this discrepancy arises from our simplifying assumption of complete spin polarization.

Examination of the electron density (not shown here) for the highest occupied Kohn-Sham wavefunction of
typical samples reveals the following.
For low concentrations $\sim 0.6 \times 10^{-7} \text{Bohr}^{-3} $ the Kohn-Sham wavefunctions resemble molecular orbitals
on clusters of two or three impurities.
As the concentration is increased towards $\sim 1.4 \times 10^{-7}  \text{Bohr}^{-3} $
the Kohn-Sham wavefunctions spread out over more impurities but remain localized.
For higher concentration $\sim 1.8 \times 10^{-7}  \text{Bohr}^{-3} $
the Kohn-Sham wavefunctions are extended across the entire sample.

\section{Conclusion}

We have presented a model for a doped semiconductor that includes both the random spatial distribution of
the donor impurities and the interaction between the carrier electrons,
and whose numerical analysis allows the metal-insulator transition to be studied.
Preliminary results indicate that the model does exhibit a metal-insulator transition.
Since in Eq. (\ref{eq:hamiltonian}) the properties of the semi-conductor enter only through the
effective mass and relative dielectric constant, we automatically recover the behavior
described in Fig. 1 of Edwards and Sienko\cite{Edwards1978} that to a good approximation the critical
concentrations in various semiconductors obey
\begin{equation}
    n_c^{1/3} a_0^*= \text{ constant}\;.
\end{equation}
However, our preliminary estimate of this constant, $\approx 0.19$, differs significantly from the observed value of
$\approx 0.26$.
As mentioned already, it is possible that this discrepancy will be resolved when our calculations
are extended to properly take account of the electron spin.

In future work we hope to shed light on the relative importance of the roles of disorder and interaction in this
transition by making a careful comparison of the critical properties of this model
with the known properties of the Anderson transition in Anderson's model of localization
for non-interacting electrons.
Due to space limitations we also defer a discussion of previous work\cite{Rose80,Bhatt81,Nielsen08}.

\section*{Acknowledgments}
This work was supported in part by Global COE Program (Core Research and Engineering of Advanced Materials-Interdisciplinary Education Center for Materials Science), MEXT, Japan.
We thank Hisazumi Akai for fruitful discussion.

\bibliographystyle{prsty2}
\bibliography{references}

\end{document}